# The Study of Jet Formation Mechanism in Fermi Blazars

Shangchun Xie[1], Zhihao Ouyang[1], Jingyu Wu[1], Hubing Xiao[1], Shaohua Zhang[1], Yongyun Chen[2], Zhijian Luo[1], and Junhui Fan[3,4,5,6]
[1] Shanghai Key Lab for Astrophysics, Shanghai Normal University, Shanghai 200234, People's Republic of China; hubing.xiao@shnu.edu.cn, zhangshaohua@shnu.edu.cn
[2] College of Physics and Electronic Engineering, Qujing Normal University, Qujing 655011, People's Republic of China
[3] Center for Astrophysics, Guangzhou University, Guangzhou 510006, People's Republic of China
[4] Great Bay Brand Center of the National Astronomical Data Center, Guangzhou 510006, People's Republic of China
[5] Key Laboratory for Astronomical Observation and Technology of Guangzhou, Guangzhou 510006, People's Republic of China
[6] Astronomy Science and Technology Research Laboratory of Department of Education of Guangdong Province, Guangzhou 510006, People's Republic of China
Received 2024 July 8; revised 2024 September 26; accepted 2024 September 29; published 2024 November 14

## Abstract

The origin of jet launching mainly comes from two mechanisms: the Blandford–Znajek (BZ) mechanism and the Blandford–Payne (BP) mechanism. However, it is in debate which one is dominating in blazars. In this work, we used a sample of 937 Fermi blazars to study the jet formation mechanism. We studied the correlation between the jet power and the accretion rate, as well as the comparison between jet power estimated by spectral energy distribution (SED) fitting and that estimated by theoretical formula and radio flux density. Our results suggest that there is no correlation between jet power estimated by SED fitting and the accretion rate for BL Lacertaes (BL Lacs), while a positive and weak correlation exists for flat spectrum radio quasars (FSRQs). Meanwhile, to confirm whether the BP and BZ mechanism is sufficient to launch the jet for FSRQs and BL Lacs, we compare the theoretical jet power with that estimated by SED fitting, as well as that by radio emission. We found that the jet power for most of the two subclasses estimated by SED fitting cannot be explained by either the BP or BZ mechanism. While the jet power for most FSRQs estimated by radio flux density can be explained by the BP mechanism, and most BL Lacs can be explained by the BZ mechanism. We also found that FSRQs have higher accretion rates than BL Lacs, implying different accretion disks around their central black holes: FSRQs typically have standard disks, while BL Lacs usually have advection-dominated accretion flow disks.

*Unified Astronomy Thesaurus concepts:* Blazars (164); Flat-spectrum radio quasars (2163); Active galactic nuclei (16); Jets (870)

*Materials only available in the online version of record:* machine-readable tables

## 1. Introduction

As an extreme subclass of Active Galactic Nuclei (AGN), blazars show some special properties, such as strong $\gamma$-ray radiation, high polarization, and rapid, drastic variability. The strong beaming effect is also a distinctive feature of blazars, which can be attributed to the small angle between the jet axis and our line of sight (C. M. Urry & P. Padovani 1995; J.-H. Fan 2002; J.-H. Fan et al. 2014; M. Lyutikov & E. V. Kravchenko 2017; Z. Ouyang et al. 2021; H. Xiao et al. 2022). Consequently, their emission is significantly Doppler boosted. According to their optical spectrum, blazars can be divided into two subclasses, i.e., flat spectrum radio quasar (FSRQ) and BL Lacertae (BL Lac), the former have strong and broad emission lines with equivalent width (EW) $\geqslant 5$ Å, while the latter have no or very weak emission lines with EW $< 5$ Å (M. Stickel et al. 1991; C. M. Urry & P. Padovani 1995; R. Scarpa & R. Falomo 1997). Also, on the basis of their synchrotron peak frequency, blazars can be classified into low-synchrotron-peaked blazars, intermediate-synchrotron-peaked blazars, and high-synchrotron-peaked blazars (E. Nieppola et al. 2006; A. A. Abdo et al. 2010; J. H. Fan et al. 2016). Typically, there is a bimodal structure in the spectral energy distribution (SED) of blazars. The bump with lower energy occurs in the infrared to X-ray band and is usually explained by synchrotron emission from relativistic electrons, while the higher energy bump occurs in the X-ray and $\gamma$-ray band and is interpreted by the inverse Compton mechanism in the leptonic model (F. Tavecchio et al. 1998; G. Ghisellini & F. Tavecchio 2009) or proton synchrotron radiation in the hadronic model (A. Mücke et al. 2003; M. Böttcher et al. 2009; M. Cerruti et al. 2015, 2019).

Since the radiation of blazars is dominated by the emission from the relativistic jet, the jet formation mechanism in blazars has long been a prosperous problem in astrophysics and has been studied by many authors. R. D. Blandford & R. L. Znajek (1977) proposed that the relativistic jet extracts the rotation energy of black holes, known as the Blandford–Znajek (BZ) mechanism. In this scenario, the jet power depends on the spin and mass of the black hole. Also, R. D. Blandford & D. G. Payne (1982) suggested that the jet power is extracted from the rotation energy of accretion disks, known as the Blandford–Payne (BP) mechanism. L. Zhang et al. (2022) found a moderate and strong, respectively, correlation between jet power and black hole spin for FSRQs and BL Lacs, which imply that the jet might be launched by BZ mechanism in both FSRQs and BL Lacs. J. Zhang et al. (2014) suggested that the jet in FSRQs may be dominated by the BZ mechanism whereas BL Lacs may be dominated by the BZ and/or the BP mechanism. L. Chen (2018) suggested that the jet formation in FSRQs is dominated by the BZ mechanism, given that jet

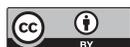

Table 1
The Sample of Fermi Blazars

| 4FGL Name (1) | z (2) | Classification (3) | $\log M_{BH}/M_\odot$ (4) | $\log L_{disk}$ (5) | $\log P_{jet}^{SED}$ (6) | References (7) | $S_\nu^{15GHz}$ (8) | References (9) |
|---|---|---|---|---|---|---|---|---|
| J0001.5+2113 | 0.439 | F | 7.54 | 44.65 | ⋯ | ⋯ | ⋯ | ⋯ |
| J0004.3+4614 | 1.81 | F | 8.36 | 46.07 | 47.38 | P17 | 0.18, 0.174 | R11, R14 |
| J0004.4−4737 | 0.88 | F | 8.28 | 45.1 | 45.88, 46.35 | G14; C23 | ⋯ | ⋯ |
| J0010.6+2043 | 0.598 | F | 7.86 | 45.35 | 48.11 | C23 | 0.08, 0.096 | R11, R14 |
| J0011.4+0057 | 1.491 | F | 8.66 | 45.71 | 47.28, 46.86 | G14; P17 | 0.299, 0.3648, 0.2493, 0.241 | R11, A11, A11, R14 |

**Note.** Column (1): source name; column (2): redshift; column (3): classification, "B" for BL Lac, "F" for FSRQ; column (4): the logarithm of black hole mass, in units of the solar mass; column (5): the logarithm of the accretion disk luminosity, in units of erg s$^{-1}$; column (6): the logarithm of jet power, in units of erg s$^{-1}$; column (7): reference of jet power (P17, V. S. Paliya et al. (2017); G14, G. Ghisellini et al. (2014), C23, Y. Chen et al. (2023c), N12, R. S. Nemmen et al. (2012); T20, C. Tan et al. (2020); C21, Y. Chen et al. (2021)); column (8): radio flux density of 15 GHz in units of Jy; column (9): reference of radio flux density (R11, J. L. Richards et al. (2011); R14, J. L. Richards et al. (2014); A11, M. Ackermann et al. (2011); A10, A. A. Abdo et al. (2010); O78, S. L. O'Dell et al. (1978); L11, M. L. Lister et al. (2011)).

(This table is available in its entirety in machine-readable form in the online article.)

power is larger than the accretion power, even L. Foschini (2011) suggested otherwise. Recently, H. Xiao et al. (2022) suggested that the jet formation in FSRQ might be dominated by the BP mechanism while BL Lacs are likely powered through the BZ mechanism.

The study of the blazar jet formation remains ongoing, and the nature of blazar jet formation stays uncertain. In this work, we use a large sample of Fermi blazars to study the jet formation. This paper is organized as follows: Section 2 describes our samples and jet models. The results are given in Section 3. The discussion is given in Section 4. Conclusions are given in Section 5. The flat lambda cold dark matter cosmology with $H_0 = 70$ km s$^{-1}$ Mpc$^{-1}$, $\Omega_M = 0.27$, and $\Omega_\Lambda = 0.73$ is applied in this work.

## 2. The Sample and the Jet Model

### 2.1. The Sample

We collected a sample of 937 Fermi blazars, including 571 FSRQs and 366 BL Lacs, with available black hole mass ($M_{BH}$) and accretion ratio ($L_{disk}/L_{Edd}$) from Y. Chen et al. (2021) and V. S. Paliya et al. (2021). We listed these sources in Table 1, columns (1), (2), and (3) give the Fermi Gamma-ray LAT DR4 (4FGL) name, redshift, and classification; The black hole mass is listed in column (4), and the accretion disk luminosity is listed in column (5). Columns (6) and (7) give the jet power ($P_{jet}^{SED}$), which is estimated based on "one-zone" leptonic model and quasi-simultaneous observation, and their corresponding references. Columns (8) and (9) give the 15 GHz flux density and their corresponding references.

### 2.2. The Jet Model

The two main jet formation mechanisms are the BZ mechanism (R. D. Blandford & R. L. Znajek 1977), in which the jet power comes from the rotation energy of the black hole, and the BP mechanism (R. D. Blandford & D. G. Payne 1982), in which the jet power comes from the accretion disk.

The maximal jet power of the BP mechanism is given by

$$P_{BP} = 4\pi \int \frac{B_{pd}^2}{4\pi} R^2 \Omega(R)\, dR \qquad (1)$$

where $B_{pd}$ is the strength of the large-scale magnetic field on the accretion disk surface, $R$ is the radius of the accretion disk, and $\Omega(R)$ is the angular velocity of the disk (P. Ghosh & M. A. Abramowicz 1997; M. Livio et al. 1999). M. Livio et al. (1999) suggested that the magnetic field can be calculated by

$$B_{pd} = \frac{H}{R} B_{dynamo}, \qquad (2)$$

where the dimensionless scale height, $\frac{H}{R}$, is defined in A. Laor & H. Netzer (1989) as

$$\frac{H}{R} = 15\dot{m}r^{-1}c_2 \qquad (3)$$

where $\dot{m} = \frac{\dot{M}}{\dot{M}_{Edd}}$, $\dot{M}_{Edd} = \frac{L_{Edd}}{\eta_{eff}c^2} = 1.4 \times 10^{18} m$ g s$^{-1}$, $m = \frac{M_{BH}}{M_\odot}$, $r = \frac{R}{R_G}$, and $R_G = \frac{GM_{BH}}{c^2}$. The variable $c_2$ is defined in Equation (5.9.10) of I. D. Novikov & K. S. Thorne (1973) as

$$c_2 = \mathscr{A}^2 \mathscr{B}^{-3} \mathscr{C}^{\frac{1}{2}} \mathscr{D}^{-1} \mathscr{E}^{-1} \mathscr{Q} \qquad (4)$$

where $\mathscr{A}$, $\mathscr{B}$, $\mathscr{C}$, $\mathscr{D}$, $\mathscr{E}$, and $\mathscr{Q}$ are the general relativistic correction factors defined in Equation (5.4.1) of I. D. Novikov & K. S. Thorne (1973) and are expressed as explicit and algebraic function of radius, except $\mathscr{Q}$. In D. N. Page & K. S. Thorne (1974), they rewrote $\mathscr{Q}$ also as an explicit and algebraic function of radius in their Equation (35) of their by combining their results with I. D. Novikov & K. S. Thorne (1973). $B_{dynamo}$ is the dynamo magnetic field strength and is given by (X. Cao 2003)

$$B_{dynamo} = 3.56 \times 10^8 r^{-\frac{3}{4}} m^{-\frac{1}{2}} \mathscr{A}^{-1} \mathscr{B} \mathscr{E}^{\frac{1}{2}} \text{ Gauss.} \qquad (5)$$

where $\mathscr{A} = 1 + j^2 x^{-4} + 2j^2 x^{-6}$, $\mathscr{B} = 1 + jx^{-3}$, $\mathscr{E} = 1 + 4j^2 x^{-4} - 4j^2 x^{-6} + 3j^4 x^{-8}$. $x = \sqrt{r}$.

In the standard accretion disk model, the angular velocity of the gas and dust in the disk is approximate to Keplerian velocity and is given by (X. Cao 2003)

$$\Omega(r) = \frac{2.03 \times 10^5}{m(r^{\frac{3}{2}} + j)} s^{-1}, \qquad (6)$$

where $j$ is the dimensionless black hole spin (X. Cao 2003). Thus, we can obtain the maximal BP jet power by inserting all the quantities above into Equation (1)

$$P_{BP} = 5.79 \times 10^{24} \frac{\dot{m}^2}{m^2} \int \frac{c_2^2 \mathscr{A}^{-2} \mathscr{B}^2 \mathscr{E}}{r^{\frac{7}{2}}(r^{\frac{3}{2}} + j)} R^2\, dR. \qquad (7)$$

In our calculation, the inner and outer radii of the accretion disk are assumed as $R_G$ and $500\,R_G$, respectively.

The maximal jet power of BZ mechanism is (D. MacDonald & K. S. Thorne 1982; P. Ghosh & M. A. Abramowicz 1997; X. Cao 2003)

$$P_{BZ} = \frac{1}{32}\omega_F^2 B_\perp^2 R_H^2 j^2 c, \quad (8)$$

where $R_H$ is the horizon radius, $R_H = \frac{2GM_{BH}}{c^2}$, $\omega_F^2 = \Omega_F(\Omega_H - \Omega_F)/\Omega_H^2$ is the measure of the effects of the angular velocity $\Omega_F$ of the field lines relative to $\Omega_H$. To maximize the power output, $\Omega_F = \frac{1}{2}\Omega_H$ is expected, and we can get $\omega_F^2 = \frac{1}{2}$. $B_\perp = \mu\kappa\sqrt{8\pi p_{max}}$ (P. Ghosh & M. A. Abramowicz 1997), where $\mu = 1.95$, $\kappa^2$ is in the range from $4 \times 10^{-3}$ to $7 \times 10^{-2}$. The typical value of $\kappa^2$ is a few percent and $\kappa^2 = 0.02$ is adopted (A. Brandenburg et al. 1996; J. F. Hawley et al. 1996; P. Ghosh & M. A. Abramowicz 1997); $p_{max} = 8.3 \times 10^5 \dot{m}_{-4}^{4/5}\chi^{-9/10}M_8^{-9/10}$ dyne cm$^{-2}$ for the gas-pressure dominated (GPD) case and $p_{max} = 1.4 \times 10^7 \alpha^{-1}M_8^{-1}$ dyne cm$^{-2}$ for radiation pressure dominated (RPD) case (I. D. Novikov & K. S. Thorne 1973; N. I. Shakura & R. A. Sunyaev 1973). $\dot{m}_{-4}$ is the accretion rate $\dot{m}$ in the units of $10^{-4}$, $\alpha$ is the viscosity parameter, $\chi = \alpha_0 + \alpha_B\kappa^2$, $\alpha_0$ is neglected, and $\alpha_B = 1.4$ is adopted (A. Brandenburg et al. 1996). So the maximal jet power of the BZ mechanism is (P. Ghosh & M. A. Abramowicz 1997)

$$P_{BZ} = \begin{cases} 8 \times 10^{42} M_8^{11/10} \dot{m}_{-4}^{4/5} j^2 & \text{GPD} \\ 2 \times 10^{44} M_8 j^2 & \text{RPD.} \end{cases} \quad (9)$$

Thus, we can obtain the maximal jet power of BP and BZ mechanism via the formulas above as long as the black hole mass ($M_{BH}$), the spin of black hole ($j$) and the accretion rate ($\dot{m}$) are known. The calculated BZ and BP jet power is listed in Table 3.

## 3. Results

### 3.1. The Correlation Between Jet Power and Accretion Ratio

To study the relationship between accretion and jet, we performed the correlation analysis between the jet power ($P_{jet}^{SED}$)[7] and the accretion ratio ($L_{disk}/L_{Edd}$) for blazars, with both available $P_{jet}^{SED}$ and $L_{disk}/L_{Edd}$, in our sample, where the Eddington luminosity $L_{Edd} = 1.3 \times 10^{38}(M_{BH}/M_\odot)$ erg s$^{-1}$. The results are shown in Figure 1 and the linear regressions obtained via the ordinary least-squares regression are

$$\log P_{jet}^{SED} = (0.21 \pm 0.04)\log\frac{L_{disk}}{L_{Edd}} + (46.96 \pm 0.06),$$

which are obtained through least square root analysis, the correlation coefficient $r = 0.25$ and chance probability $p = 5.1 \times 10^{-9}$ obtained via Pearson analysis; and

$$\log P_{jet}^{SED} = (0.16 \pm 0.07)\log\frac{L_{disk}}{L_{Edd}} + (46.96 \pm 0.07),$$

---

[7] We note that in V. S. Paliya et al. (2017), $P_{jet} = P_{ele} + P_{mag} + P_{kin}$, i.e., the sum of jet power of the relativistic electrons, magnetic field, and cold protons. But we also included that of radiation $P_{rad}$.

with $r = 0.11$ and $p = 0.02$ for FSRQs in this work;

$$\log P_{jet}^{SED} = (0.01 \pm 0.09)\log\frac{L_{disk}}{L_{Edd}} + (46.29 \pm 0.26),$$

with $r = 0.008$ and $p = 0.92$ for BL Lacs in this work.

The $p$-value is larger than the threshold of 0.05 for BL Lacs, implying the relationship is not statistically significant and may not related to the accretion rate or process. While there is a positive and weak relationship for FSRQs. This result indicates that accretion may play a more important role in jet launching for FSRQs than for BL Lacs.

### 3.2. The Jet Power Under the BZ and BP Mechanisms

The advection-dominated accretion flow (ADAF) disk and the standard disk are believed to be GPD and RPD, respectively (R. Narayan et al. 1998; L. Foschini 2011). As we will show in the next section, the FSRQ is believed to hold the standard disk and the BL Lac is believed to hold the ADAF disk. Thus, we considered the BZ jet power for FSRQs under the RPD condition and the BZ jet power for BL Lacs only under the GPD condition, the black hole spin $j = 0.95$ is adopted in our calculation. The comparison results between the observed jet power ($P_{jet}^{SED}$) and the calculated theoretical maximal jet power ($P_{jet}^{BZ}$ and $P_{jet}^{BP}$) are shown in Figure 2.

In the upper panel of Figure 2, there are 395 of 399 FSRQs and 146 of 162 BL Lacs that lie below the equality line, indicating that the BZ mechanism may not be sufficient for either FSRQs or BL Lacs. In the bottom panel, 192 of 399 FSRQs and 153 of 162 BL Lacs lie below the equality line, suggesting that the BP mechanism is able to maintain the jet power for about half of the FSRQs and only a few of the BL Lacs.

### 3.3. The Property of the Blazar Accretion Disk

The line dimensionless accretion rate can be expressed as

$$\lambda = \frac{L_{lines}}{L_{Edd}}, \quad L_{lines} = \xi L_{disk}, \quad (10)$$

(J.-M. Wang et al. 2002), where $\xi = 0.1$ (H. Netzer 1990), and $L_{lines}$ is the total luminosity of emission lines. For an optically thin ADAF disk, the relation between $\lambda$ and the dimensionless accretion rate $\dot{m}$, is given by J.-M. Wang et al. (2002)

$$\dot{m} = 2.17 \times 10^{-2} \alpha_{0.3}\xi_{-1}^{-\frac{1}{2}}\lambda_{-4}^{\frac{1}{2}} \quad (11)$$

where the assumption that the total disk luminosity from the ADAF disk is $L_{disk} \propto \alpha^{-2}M_{BH}\dot{m}^2$ (R. Mahadevan 1997), $\alpha$ is the viscosity parameter, and $\alpha_{0.3} = \alpha/0.3$, $\xi_{-1} = \xi/0.1$, $\lambda_{-4} = \lambda/10^{-4}$. The presence of an ADAF disk requires $\dot{m} \leqslant \alpha^2$ (R. Narayan et al. 1998). Thus, Equation (11) can be expressed as

$$\lambda_l = 1.72 \times 10^{-3}\xi_{-1}\alpha_{0.3}^2. \quad (12)$$

Thus, an optically thin ADAF disk requires $\lambda < \lambda_l$ and $\alpha = 0.3$ is adopted.

Optically thick and geometrically thin standard disk requires

$$\dot{m} = \frac{L_{lines}}{\xi L_{Edd}} = 10\xi_{-1}^{-1}\lambda, \quad (13)$$

the necessary condition for the existence of a standard disk is $\alpha^2 \leqslant \dot{m} < 1$ (N. I. Shakura & R. A. Sunyaev 1973), which

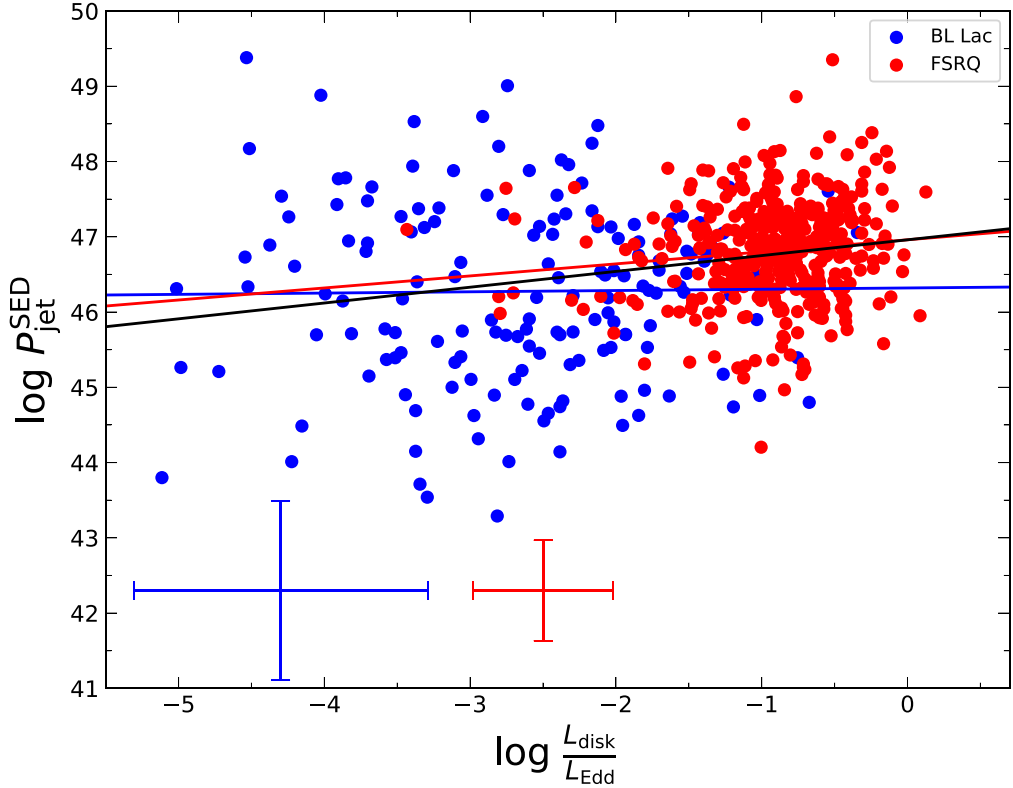

**Figure 1.** The correlation between jet power and accretion rate. The red dots and line stand for the data points and corresponding linear regression of FSRQs; the blue dots and line stand for the data points and corresponding linear regression of BL Lacs. The solid black line represents the linear regression for all blazars in this diagram. The blue and red crosses are the standard deviation error for BL Lac and FSRQ, respectively.

gives

$$\lambda_2 = 9.0 \times 10^{-3} \xi_{-1} \alpha_{0.3}^2, \quad (14)$$

the standard disk emerges when $\lambda \geqslant \lambda_2$.

When accretion rate $\dot{m} \geqslant 1$, which gives

$$\lambda_3 = 0.1 \xi_{-1}, \quad (15)$$

a super Eddington accretion (SEA) slim disk exists, and this requires $\lambda \geqslant \lambda_3$ (J.-M. Wang et al. 2002, 2003).

According to Equation (10), assuming $\xi = 0.1$, we can get $\log \lambda$ for those sources with available $\log \frac{L_{disk}}{L_{Edd}}$. The distribution of the $\log \lambda$ of sources in our sample is shown in Figure 3. The distribution of $\log \lambda$ is divided into four regions by the three boundaries stated above, corresponding to different accretion disk models. Region A represents the ADAF disk, region B is a transition from the ADAF disk to the standard disk, region C stands for the standard disk, and region D stands for the SEA slim disk. We can see that FSRQs have a different accretion disk model compared to BL Lacs. Most of FSRQs have thin disks while BL Lacs have ADAF disks. The different accretion disk model may represent the different dominated jet formation mechanisms in FSRQs and BL Lacs (G. Ghisellini & A. Celotti 2001; H.-Y. Pu et al. 2012).

## 4. Discussion

### 4.1. The Correlation Between Jet Power and Accretion Ratio

The jet formation mechanism in blazars might be extracting energy from the black hole, i.e., BZ mechanism (R. D. Blandford & R. L. Znajek 1977) or the rotation energy of the accretion disk, i.e., BP mechanism (R. D. Blandford & D. G. Payne 1982), which has been studied by many previous works.

The relation between jet power and black hole spin and mass, accretion disk luminosity and accretion rate have been studied extensively (G. Ghisellini et al. 2014; D. R. Xiong & X. Zhang 2014; T. Sbarrato et al. 2016; L. Chen 2018; L. Zhang et al. 2020, 2022; Y. Chen et al. 2021; H. Xiao et al. 2022). Y.-Y. Chen et al. (2015) proposed that the BZ mechanism may not be sufficient to generate jets in blazars. Later works, for instance, H. Xiao et al. (2022), found that the BZ mechanism is not consistent with the relationship observed between jet power and normalized accretion disk luminosity in FSRQs, suggesting that jets in FSRQs may be dominated by accretion disks. Y. Chen et al. (2023a) found that jet kinetic power, which is believed to be about 90% of the jet power, of about 72% intermediate peak frequency BL Lacs and 94% high-frequency peak BL Lacs of their sample can be explained by the hybrid model in the ADAF scenrio.

In Figure 1, we found weak positive correlations between $\log P_{jet}^{SED}$ and $\log \frac{L_{disk}}{L_{Edd}}$ for the entire blazar sample and for FSRQs in this work, while no correlation was found for BL Lacs. It is possible that the accretion disk is not the only source of charging the relativistic jet, the black hole may also play an important role in it. We employ multiple linear regression analysis to obtain the correlation between the jet power and both accretion rate and black hole mass for available FSRQs

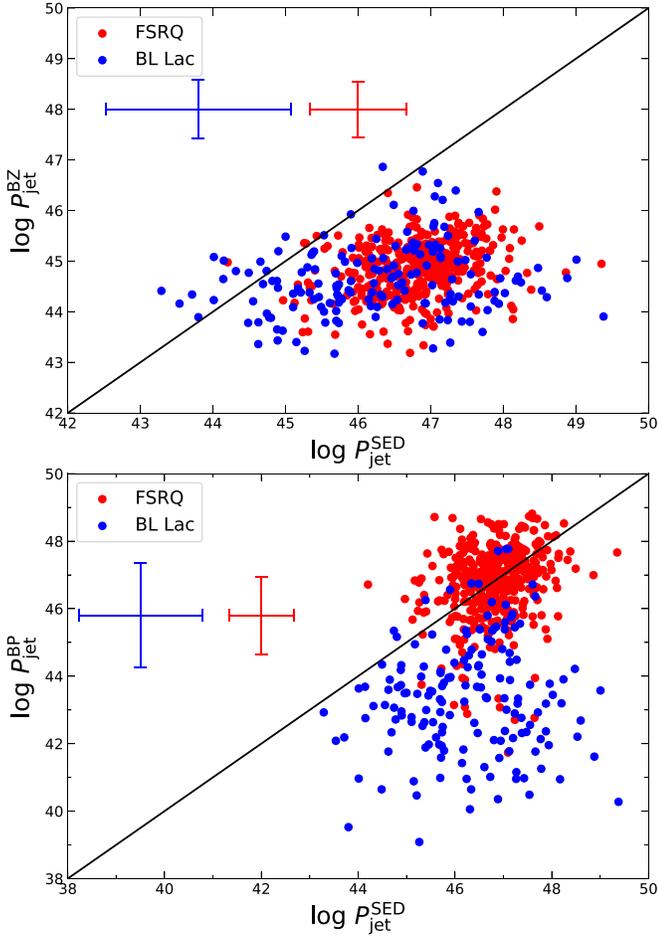

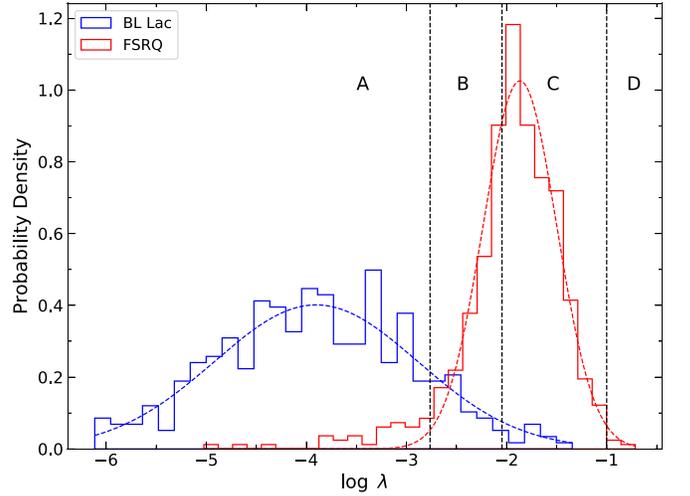

**Figure 2.** The maximal BP and BZ jet power calculated by Equations (7) and (9) vs. the observed jet power of samples. The upper panel shows the calculated theoretical maximal BZ jet power vs. the observed jet power, and the bottom panel shows the calculated theoretical maximal BP jet power vs. the observed jet power. The black lines are the equality lines. The blue and red crosses are the same as those in Figure 1.

**Figure 3.** The distribution of the log $\lambda$ of our sample. The red denotes FSRQs and the blue denotes BL Lacs. The dashed curves are the Gaussian fit of the histogram.

and BL Lacs in our sample, and the result gives

$$\log P_{\rm jet}^{\rm SED} = (0.18 \pm 0.06) \log \frac{L_{\rm disk}}{L_{\rm Edd}} + (0.33 \pm 0.06) \log M_{\rm BH} + (44.11 \pm 0.49)$$

for FSRQs with $r = 0.29$ and $p = 2.48 \times 10^{-9}$. The $F$-test, which is employed to test the significance of adding the black hole mass as a new parameter, gives $r_F = 17.74$ probability $p_F = 4.19 \times 10^{-8}$. The $p_F$ is smaller than the threshold of 0.05, and suggests we should reject the null hypothesis that the linear regression between $\log P_{\rm jet}^{\rm SED}$ and $\log \frac{L_{\rm disk}}{L_{\rm Edd}}$ in Section 3 is sufficient. A multiple linear correlation for BL Lacs gives

$$\log P_{\rm jet}^{\rm SED} = (0.21 \pm 0.12) \log \frac{L_{\rm disk}}{L_{\rm Edd}} + (0.47 \pm 0.17) \log M_{\rm BH} + (42.77 \pm 1.28)$$

with $r = 0.21$ and $p = 0.001$, and $r_F = 3.94$ and $p_F = 0.02$. Our multiple analysis results suggest that jet power depends on both the accretion rate and black hole mass for both FSRQs and BL Lacs.

S. Heinz & R. A. Sunyaev (2003) proposed the theoretical analysis of the dependence of jet power on the Eddington ratio and black hole mass. For standard accretion, the core flux $F_\nu \sim M_{\rm BH}^{\frac{17}{12}}$; for radiatively inefficient accretion, $F_\nu \propto (M_{\rm BH} \dot{m})^{\frac{17}{12}}$. Some previous literature have also studied this correlation, as listed in Table 2. In general, these results are consistent with ours that blazar jet power is dependent on both accretion ratio and black hole mass. Among these results, the slopes of $\log M_{\rm BH}$ tend to be greater than the slopes for $\log \frac{L_{\rm disk}}{L_{\rm Edd}}$ suggesting that black hole may take a more important role of fueling a jet than the accretion disk does.

### 4.2. The Mechanism of Launching the Blazar Jet

In general, neither the BZ mechanism nor the BP mechanism can fully explain the jet-powering of blazars, and the reasons are twofold. On the one hand, the jet power is likely to be constructed by both the black hole and the accretion disk, namely the BZ mechanism and the BP mechanism, as we have discussed as previous section.

On the other hand, the method of measuring the observed jet power could be inaccurate. The jet power measurements used in this work are obtained through broadband SED fitting; the broadband observation is usually performed during flaring or other active states of blazars, during which additional microphysics could take place such as magnetic reconnection, shock acceleration, and shear acceleration. These processes lead to enhanced emission and acceleration of particles. However, they are not included in the estimates of jet power. As such the jet power from SED fitting could be overestimated compared to the theoretical models. Additionally, these models are typically "one-zone" models that ignore the underlying jet structure and may underestimate the jet power compared to "multizone" models. Thus it is necessary to explore appropriate methods of estimating jet power.

Recently, L. Foschini et al. (2024) made a comparison between different methods to estimate the power of relativistic jets and presented an easy-to-use equation based on radio observations and the evergreen R. D. Blandford & A. Königl (1979) model,

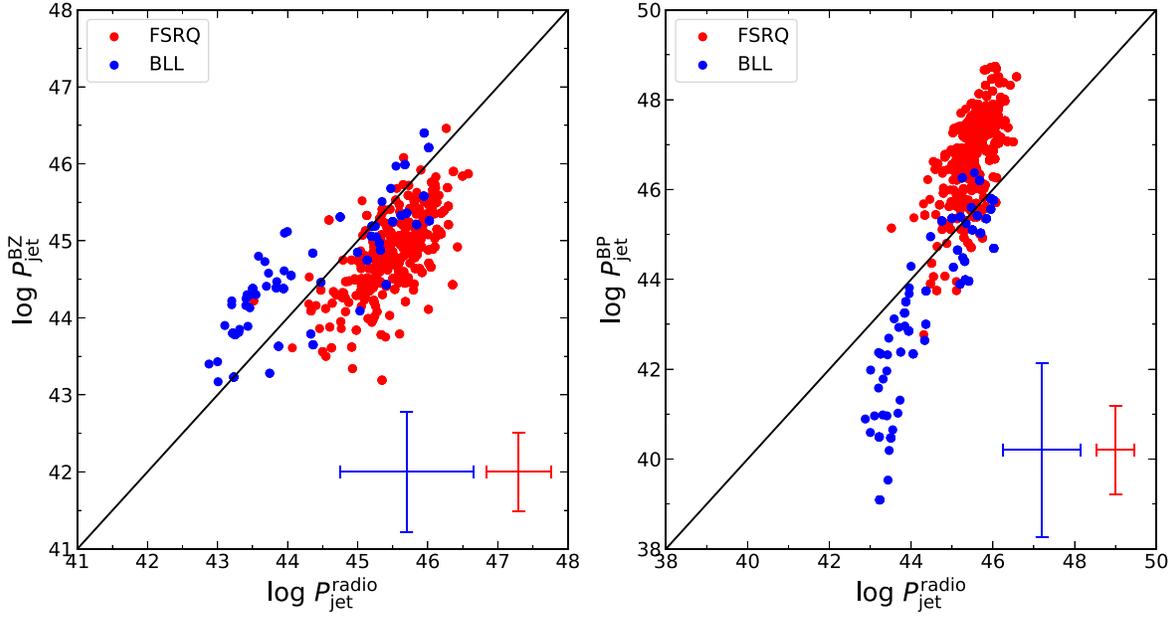

**Figure 4.** The correlation between jet power calculated by radio flux density and that by BP and BZ mechanism. The blue and red crosses are the same as those in Figure 1.

**Table 2**
Multiple Regression Results from Literature

| Multiple Regression Result | References |
| --- | --- |
| $\log P_{\rm jet} = 0.22 \log(L_{\rm bol}/L_{\rm Edd}) + 0.59 \log M_{\rm BH} + 40.48$ | Y. Liu et al. (2006) |
| $\log P_{\rm jet} = 0.52 \log(L_{\rm bol}/L_{\rm Edd}) + 0.54 \log M_{\rm BH} + 43.15$ | D. R. Xiong & X. Zhang (2014) |
| $\log P_{\rm jet} = 0.62 \log(L_{\rm bol}/L_{\rm Edd}) + 1.01 \log M_{\rm BH} + 39.77$ (FSRQs) | Y. Chen et al. (2023b) |
| $\log P_{\rm jet} = 0.99 \log(L_{\rm bol}/L_{\rm Edd}) + 0.89 \log M_{\rm BH} + 39.96$ (BL Lacs) | Y. Chen et al. (2023b) |

**Note.** We note that $L_{\rm bol}$ is the bolometric luminosity and $L_{\rm bol} = 10 L_{\rm BLR}$ in their works. The same as in V. S. Paliya et al. (2021).

the total jet power can be estimated by

$$P_{\rm jet}^{\rm radio} = (4.5 \times 10^{44}) \left( \frac{S_\nu d_{\rm L,9}^2}{1+z} \right)^{\frac{12}{17}} \quad (16)$$

where $S_\nu$ is the observed radio flux density in the units of Jy, $d_{\rm L,9}$ is luminosity distance in the units of Gpc (R. D. Blandford & A. Königl 1979; L. Foschini et al. 2024). The biases of this model mainly stem from the adaptation of median or weighted mean values calculated over long periods, despite the strong variability of jetted AGN, as well as some physical factors considered in the same model. Meanwhile, the R. D. Blandford & A. Königl (1979) model is for flat spectrum sources and may imply a large deviation for steep spectrum sources. Additionally, the model should be employed carefully when dealing with extremely weak or extremely powerful jets (L. Foschini et al. 2024). The jet power estimated by this method is regarded as an average over the source lifetime (P. Pjanka et al. 2017). We collected 15 GHz flux density in units of Jy from the literature and listed them in columns (8) and (9) of Table 1. For sources with more than one flux measurement reported, the average value is used. Then we calculated and compared the $P_{\rm jet}^{\rm radio}$ with the calculated theoretical BZ and BP jet power.

We also note that the limitation of 15 GHz data includes that the flux variability at this frequency may be due to not only the radiative losses, but may also be affected by other dissipation mechanisms. So the measurement of the Doppler factor via the brightness temperature can be biased, and thus affect the jet power (S. G. Jorstad et al. 2005; D. C. Homan et al. 2021; L. Foschini et al. 2024).

The results of the comparison between the observed radio-based jet power ($P_{\rm jet}^{\rm radio}$) and the calculated theoretical maximal jet power ($P_{\rm jet}^{\rm BZ}$ and $P_{\rm jet}^{\rm BP}$) are shown in Figure 4. In the left panel, 273 of 287 FSRQs lie below the equality line, while 14 lie above the line; 37 of 60 BL Lacs lie above the line, and 23 lie below the line. In the right panel, 264 of 287 FSRQs lie above the equality line, 23 lie below the line; 51 of 60 BL Lacs lie below the line and 9 BL Lacs lie above the line (as listed in Table 3).

The results suggest that FSRQ jets are produced by the BP mechanism, while the BZ mechanism might not be sufficient. Y. Chen et al. (2023c) also found that FSRQs' jet power can be explained by the BP mechanism, and the result is consistent with that in H. Xiao et al. (2022). For BL Lacs, the BZ mechanism may be sufficient for most of BL Lacs to launch jets, and the results is agreement with that in H. Xiao et al. (2022).

### 4.3. The Accretion Disk

As we have demonstrated the accretion property in Figure 3, we find that most BL Lacs have ADAF disks, while most of the

**Table 3**
The Estimated Jet Power Through Equations (7) and (9); the Comparison Between SED and Radio-based Jet Power, and the Theoretical Jet Power

| 4FGL Name | log $P_{\rm jet}^{\rm BZ}$ | log $P_{\rm jet}^{\rm BP}$ | BZ Mechanism | | BP Mechanism | |
|---|---|---|---|---|---|---|
| | | | log $P_{\rm jet}^{\rm SED}$ | log $P_{\rm jet}^{\rm radio}$ | log $P_{\rm jet}^{\rm SED}$ | log $P_{\rm jet}^{\rm radio}$ |
| J0001.5+2113 | 43.80 | 45.54 | ⋯ | ⋯ | ⋯ | ⋯ |
| J0004.3+4614 | 44.62 | 47.56 | N | N | Y | ⋯ |
| J0004.4−4737 | 44.54 | 45.70 | N | ⋯ | N | ⋯ |
| J0010.6+2043 | 44.12 | 46.62 | N | N | N | Y |
| J0011.4+0057 | 44.92 | 46.54 | N | N | N | Y |

**Note.** If the jet power can be explained by the BZ mechanism, "Y" stands for yes, "N" stands for no; if the jet power can be explained by the BP mechanism, "Y" stands for yes, "N" stands for no.

(This table is available in its entirety in machine-readable form in the online article.)

FSRQs have standard disks. Previous studies have suggested that the transition from ADAF disks to standard disks is possible (F. Meyer & E. Meyer-Hofmeister 1994; R. Narayan 1996; A. A. Esin et al. 1997; J.-F. Lu et al. 2004; F. Yuan & R. Narayan 2004; C. Done et al. 2007; R. Narayan & J. E. McClintock 2008). FSRQs are believed to be in a gas-rich environment and have strong emission lines, while BL Lacs are situated in a gas-poor environment and have no or weak emission lines. As the gas is gradually exhausted, the accretion rate and radiation efficiency also decrease, then the transition from the standard disk to a combined disk occurs, which is characterized by an ADAF inner region (at least at small radii) and a standard disk outer region. When the accretion rate continues to decrease, the transition radius continues to increase until it disappears, and the accretion disk eventually becomes an ADAF disk.

From Figure 3, we can see clearly different distributions: FSRQs have a higher accretion rate and narrower distributions than BL Lacs. The Gaussian fit was applied to Figure 3 and the fitting results give a mean value log $\lambda = -3.90$ with a standard deviation of 1.02 for BL Lacs and a mean value log $\lambda = -1.87$ with the standard deviation of 0.37 for FSRQs. We also suggest using the Gaussian intersection point value, log $\lambda = -2.57$, corresponding to the accretion rate log $\frac{L_{\rm disk}}{L_{\rm Edd}} = -1.57$, of the two Gaussian profiles to divide FSRQs and BL Lacs. G. Ghisellini et al. (2010) suggested that the division between BL Lacs and FSRQs occurs at $L_{\rm disk}/L_{\rm Edd} \sim 0.01$, which is approximate to the division between FR I and FR II galaxies (G. Ghisellini & A. Celotti 2001). Similarly, T. Sbarrato et al. (2012) also found that the division between BL Lacs and FSRQs emerges at about $L_{\rm disk}/L_{\rm Edd} \sim 5 \times 10^{-3}$–$10^{-2}$, and the division might be due to the transition from radiatively inefficient to efficient disks. Our result is roughly consistent with theirs.

The radial velocity of geometrically thick ADAF is much larger than that of a standard disk, the magnetic field can be dragged into the vicinity of the black hole more efficiently, and amplified by ADAF disks, so the BZ mechanism could be more important than BP mechanism in ADAF disks (R. Narayan & I. Yi 1994, 1995; R. Narayan & J. E. McClintock 2008; X. Cao 2011). In contrast, the production of jets in thin disks is expected to be suppressed because of the lack of strong poloidal magnetic fields, the low accretion rate disks may provide an especially favorable environment for jet production. S. H. Lubow et al. (1994) reported that the standard, geometrically thin accretion disks cannot drag the magnetic flux to the center of the disk because of the outward diffusion of the magnetic field in the turbulence triggered by the magnetorotational instability. Similarly, M. Sikora et al. (2013) suggested a two-phase scenario, i.e., a cold, thin accretion phase following a lower accretion rate, with a hot accretion phase taking place in radio-loud sources. The large magnetic fluxes were accumulated during the hot accretion phase.

Moreover, the gas in ADAF disks are believed to have a positive Bernoulli parameter, which is the sum of the kinetic energy, potential energy, and enthalpy, from which one can infer that the gas is not bound to the black hole and strong outflows and jets are common in ADAFs (R. Narayan & I. Yi 1994, 1995; R. D. Blandford & M. C. Begelman 1999). An important consequence is that the gas accreting onto the black hole is much less than the mass supplied at the outer edge of the accretion flow (R. Narayan & J. E. McClintock 2008), i.e., the accretion rate increase with radius, $\dot{M}(R) \propto R^s$. Then the accretion power around the inner region of the disk may be much lower than that of a standard disk and not sufficient to launch a powerful jet. So other physical processes and energy might be needed to generate powerful jets.

However recent thin disk simulations have shown that a weak seed field that is confined to the disk evolves to a global magnetic field, as well as a strong field that has been advected to the central black hole and launches the BZ jet. The large global magnetic flux can drive the strong disk wind, which can be collimated to the disk jet, implying that the BP jet is also possible (I. K. Dihingia & C. Fendt 2024). Therefore, the hybrid models (i.e., BZ+BP) are highly likely and could maybe explain the current failure of the BZ-only models.

The key parameter of launching powerful jets is the magnetic flux threading the black hole horizon (R. D. Blandford & R. L. Znajek 1977; M. Sikora & M. C. Begelman 2013), the generation efficiency of jets is also dependent on the magnetic flux dragged to the center (A. Tchekhovskoy et al. 2011). Thus, a geometrically thick disk is required to transport a large amount of magnetic flux to the center (X. Cao 2011; J. C. McKinney et al. 2012), due to the vertical thickness that supports a strong poloidal magnetic field (M. Livio et al. 1999; D. L. Meier et al. 2001; R. Narayan 2012). When the magnetic flux saturates, a magnetically arrested disk (MAD) exists (R. Narayan et al. 2003; A. Tchekhovskoy et al. 2011). Considering the MAD scenario, the saturation field strength is approximately (R. Narayan et al. 2003; F. Yuan & R. Narayan 2014)

$$B_{\rm MAD} = 1.5 \times 10^9 (1-f_\Omega)^{\frac{1}{2}} \epsilon^{-\frac{1}{2}} m^{-\frac{1}{2}} \dot{m}^{\frac{1}{2}} r^{-\frac{5}{4}} \text{ Gauss}, \quad (17)$$

where $\epsilon$ is the ratio of the velocity of the gas in MAD to the freefall velocity (R. Narayan et al. 2003; R. Narayan 2012), $f_\Omega$ is the ratio of the angular velocity of the accretion flow and the freefall angular velocity, and $r$ is the radius in the units of $R_G$. We assume $\epsilon = 0.01$, $f_\Omega = 0.5$, and $r \sim R_{ISCO}$ (H. He et al. 2024)

$$R_{ISCO} = 3 + z_2 - [(3 - z_1)(3 + z_1 + 2z_2)]^{\frac{1}{2}}, \quad (18)$$

where $z_1 = 1 + (1 - j^2)^{\frac{1}{3}}[(1 + j)^{\frac{1}{3}} + (1 - j)^{\frac{1}{3}}]$, $z_2 = (3j^2 + z_1^2)^{\frac{1}{2}}$. Therefore, given the black hole mass, spin, and accretion rate, the jet power in the scenario of the MAD can be estimated by substituting Equations (17) and (18) into Equation (8). As discussed above, we employed this method only to BL Lacs. The result is shown in Figure 5.

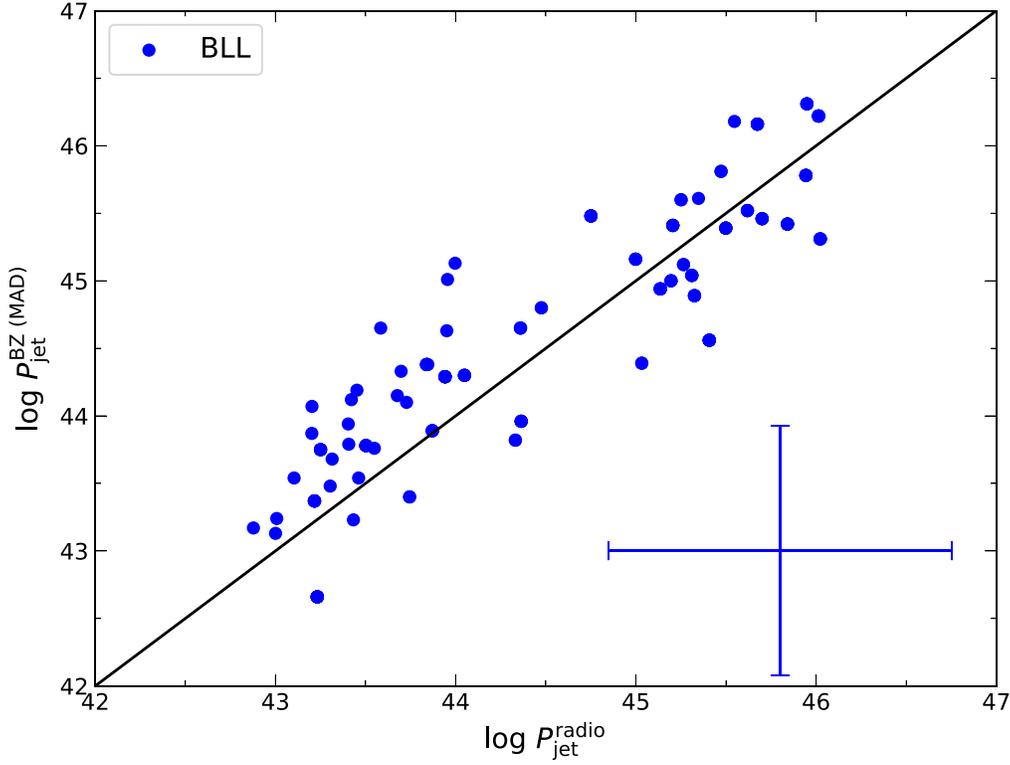

**Figure 5.** The comparison between jet power in the MAD scenario and that estimated by 15 GHz flux density for BL Lacs. The black line is the equality line. The blue and red crosses are the same as those in Figure 1.

In Figure 5, 41 of 60 BL Lacs lie above the equality line, 19 lie below the line. The MAD scenario can explain most of jet power estimated by radio emission. The results may imply the possibility of the existence of MAD in the BL Lacs's history lifetime or even the present. M. Zamaninasab et al. (2014) found that most, if not all, radio-loud sources contain dynamically important magnetic flux near the central black holes. Radio-loud AGN is conducive to the formation of MAD disks, which powers the jet via the BZ mechanism. Both our results and theirs suggest that the MAD disk may be common in blazars.

H. Yang et al. (2024) has revealed that, by comparing the results obtained from general relativistic magnetohydrodynamic (GRMHD) simulations with the millimeter observations, the jet in M87 is launched by extracting the rotation energy of a high spin black hole surrounded by a MAD disk. The GRMHD simulation strongly prefers a high-spinning black hole and MAD model, indicating that the BZ mechanism works (Event Horizon Telescope Collaboration et al. 2019).

## 5. Conclusion

In this work, to study the jet formation mechanism in blazars, we collected a sample of 937 Fermi blazars with available black hole mass, accretion disk luminosity, jet power, and radio flux density. We studied the correlations between observed jet power and the accretion rate and between the observed jet power and theoretical jet power. Our main conclusions are as follows:

1. We found no correlation between the jet power and the accretion rate for BL Lacs and a positive and weak correlation for FSRQs, which indicates different dominant jet formation mechanisms for BL Lacs and FSRQs. Multiple linear regression gives $\log P_{jet}^{SED} = (0.18 \pm 0.06)\log \frac{L_{disk}}{L_{Edd}} + (0.33 \pm 0.06)\log M_{BH} + (44.11 \pm 0.49)$ for FSRQs and $\log P_{jet}^{SED} = (0.21 \pm 0.12)\log \frac{L_{disk}}{L_{Edd}} + (0.47 \pm 0.17)\log M_{BH} + (42.77 \pm 1.28)$ for BL Lacs, which may imply that jet power comes from both accretion rate and black hole mass for FSRQs and BL Lacs.

2. Our results suggested that neither the BP mechanism nor the BZ mechanism can explain the jet power estimated by SED fitting. When we compared the jet power derived from the radio flux density, which usually presents a time-averaged value across the blazar's lifetime, with the theoretical jet power, the result suggests that most of FSRQs can be explained by the BP mechanism, while most of BL Lacs can be explained by the BZ mechanism.

3. We also found that FSRQs, with an average accretion rate of $\log \lambda = -3.90$, have higher accretion rates than BL Lacs, with an average accretion rate of $\log \lambda = -1.87$, implying different accretion disks around their central black holes: FSRQs typically have standard disks, while BL Lacs usually have ADAF disks. By comparing the jet power estimated from radio emission with that in the MAD scenario, we found that most BL Lacs can be explained by the MAD scenario. This suggests the possibility of the existence of MAD in the history or even the present state of blazars, especially for BL Lacs.

### Acknowledgments

We thank the support from our laboratory, the Key Laboratory for Astrophysics of Shanghai. H.B.X. acknowledges the support from the National Natural Science Foundation of China (NSFC) under grant No. 12203034, from the Shanghai Science and Technology Fund under grant No. 22YF1431500, and from the science research grants from the China Manned Space Project. S.H.Z. acknowledges support from the National Natural Science Foundation of China (grant No. 12173026), the National Key Research and Development Program of China (grant No. 2022YFC2807303), the Shanghai Science and Technology Fund (grant No. 23010503900), the Program for Professor of Special Appointment (Eastern Scholar) at Shanghai Institutions of Higher Learning and the Shuguang Program (23SG39) of the Shanghai Education Development Foundation and Shanghai Municipal Education Commission. Z.J.L. acknowledges the support from NSFC grant 12141302 and the science research grants from the China Manned Space Project. J.H.F. acknowledges the support of the NSFC U2031201, NSFC 11733001, NSFC 12433004, the Scientific and Technological Cooperation Projects (20202023) between the People's Republic of China and the Republic of Bulgaria, the science research grants from the China Manned Space Project with No. CMS-CSST-2021-A06.

### ORCID iDs

Hubing Xiao https://orcid.org/0000-0001-8244-1229
Shaohua Zhang https://orcid.org/0000-0001-8485-2814
Yongyun Chen https://orcid.org/0000-0001-5895-0189
Junhui Fan https://orcid.org/0000-0002-5929-0968

### References

Abdo, A. A., Ackermann, M., Agudo, I., et al. 2010, ApJ, 716, 30
Ackermann, M., Ajello, M., Allafort, A., et al. 2011, ApJ, 741, 30
Blandford, R. D., & Begelman, M. C. 1999, MNRAS, 303, L1
Blandford, R. D., & Königl, A. 1979, ApJ, 232, 34
Blandford, R. D., & Payne, D. G. 1982, MNRAS, 199, 883
Blandford, R. D., & Znajek, R. L. 1977, MNRAS, 179, 433
Böttcher, M., Reimer, A., & Marscher, A. P. 2009, ApJ, 703, 1168
Brandenburg, A., Nordlund, A., Stein, R. F., & Torkelsson, U. 1996, ApJL, 458, L45
Cao, X. 2003, ApJ, 599, 147
Cao, X. 2011, ApJ, 737, 94
Cerruti, M., Zech, A., Boisson, C., & Inoue, S. 2015, MNRAS, 448, 910
Cerruti, M., Zech, A., Boisson, C., et al. 2019, MNRAS, 483, L12
Chen, L. 2018, ApJS, 235, 39
Chen, Y., Gu, Q., Fan, J., et al. 2021, ApJ, 913, 93
Chen, Y., Gu, Q., Fan, J., et al. 2023a, MNRAS, 526, 4079
Chen, Y., Gu, Q., Fan, J., et al. 2023b, MNRAS, 519, 6199
Chen, Y., Gu, Q., Fan, J., et al. 2023c, ApJ, 268, 10
Chen, Y.-Y., Zhang, X., Xiong, D., & Yu, X. 2015, AJ, 150, 8
Dihingia, I. K., & Fendt, C. 2024, arXiv:2404.06140
Done, C., Gierliński, M., & Kubota, A. 2007, A&ARv, 15, 1
Esin, A. A., McClintock, J. E., & Narayan, R. 1997, ApJ, 489, 865
Event Horizon Telescope Collaboration, Akiyama, K., Alberdi, A., et al. 2019, ApJL, 875, L5
Fan, J.-H. 2002, PASJ, 54, L55
Fan, J.-H., Bastieri, D., Yang, J.-H., et al. 2014, RAA, 14, 1135
Fan, J. H., Yang, J. H., Liu, Y., et al. 2016, ApJS, 226, 20
Foschini, L. 2011, RAA, 11, 1266
Foschini, L., Dalla Barba, B., Tornikoski, M., et al. 2024, Univ, 10, 156
Ghisellini, G., & Celotti, A. 2001, A&A, 379, L1
Ghisellini, G., & Tavecchio, F. 2009, MNRAS, 397, 985
Ghisellini, G., Tavecchio, F., Foschini, L., et al. 2010, MNRAS, 402, 497
Ghisellini, G., Tavecchio, F., Maraschi, L., Celotti, A., & Sbarrato, T. 2014, Natur, 515, 376
Ghosh, P., & Abramowicz, M. A. 1997, MNRAS, 292, 887
Hawley, J. F., Gammie, C. F., & Balbus, S. A. 1996, ApJ, 464, 690
He, H., You, B., Jiang, N., et al. 2024, MNRAS, 530, 530
Heinz, S., & Sunyaev, R. A. 2003, MNRAS, 343, L59
Homan, D. C., Cohen, M. H., Hovatta, T., et al. 2021, ApJ, 923, 67
Jorstad, S. G., Marscher, A. P., Lister, M. L., et al. 2005, AJ, 130, 1418
Laor, A., & Netzer, H. 1989, MNRAS, 238, 897
Lister, M. L., Aller, M., Aller, H., et al. 2011, ApJ, 742, 27
Liu, Y., Jiang, D. R., & Gu, M. F. 2006, ApJ, 637, 669
Livio, M., Ogilvie, G. I., & Pringle, J. E. 1999, ApJ, 512, 100
Lu, J.-F., Lin, Y.-Q., & Gu, W.-M. 2004, ApJL, 602, L37
Lubow, S. H., Papaloizou, J. C. B., & Pringle, J. E. 1994, MNRAS, 267, 235
Lyutikov, M., & Kravchenko, E. V. 2017, MNRAS, 467, 3876
MacDonald, D., & Thorne, K. S. 1982, MNRAS, 198, 345
Mahadevan, R. 1997, ApJ, 477, 585
McKinney, J. C., Tchekhovskoy, A., & Blandford, R. D. 2012, MNRAS, 423, 3083
Meier, D. L., Koide, S., & Uchida, Y. 2001, Sci, 291, 84
Meyer, F., & Meyer-Hofmeister, E. 1994, A&A, 288, 175
Mücke, A., Protheroe, R. J., Engel, R., Rachen, J. P., & Stanev, T. 2003, APh, 18, 593
Narayan, R. 1996, ApJ, 462, 136
Narayan, R., Igumenshchev, I. V., & Abramowicz, M. A. 2003, PASJ, 55, L69
Narayan, R., Mahadevan, R., & Quataert, E. 1998, in Theory of Black Hole Accretion Disks, ed. M. A. Abramowicz, G. Björnsson, & J. E. Pringle (Cambridge: Cambridge Univ. Press), 148
Narayan, R., & McClintock, J. E. 2008, NewAR, 51, 733
Narayan, R., SÄ dowski, A., Penna, R. F, & Kulkarni, A. K. 2012, MNRAS, 426, 3241
Narayan, R., & Yi, I. 1994, ApJL, 428, L13
Narayan, R., & Yi, I. 1995, ApJ, 452, 710
Nemmen, R. S., Georganopoulos, M., Guiriec, S., et al. 2012, Sci, 338, 1445
Netzer, H. 1990, in Active Galactic Nuclei, ed. T. J-L. Courvoisier & M. Mayor (Berlin: Springer), 57
Nieppola, E., Tornikoski, M., & Valtaoja, E. 2006, A&A, 445, 441
Novikov, I. D., & Thorne, K. S. 1973, in Black Holes (Les Astres Occlus) (New York: Gordon & Breach), 343
O'Dell, S. L., Puschell, J. J., Stein, W. A., et al. 1978, ApJ, 224, 22
Ouyang, Z., Xiao, H., Zheng, Y., Xu, P., & Fan, J. 2021, Ap&SS, 366, 12
Page, D. N., & Thorne, K. S. 1974, ApJ, 191, 499
Paliya, V. S., Domínguez, A., Ajello, M., Olmo-García, A., & Hartmann, D. 2021, ApJS, 253, 46
Paliya, V. S., Marcotulli, L., Ajello, M., et al. 2017, ApJ, 851, 33
Pjanka, P., Zdziarski, A. A., & Sikora, M. 2017, MNRAS, 465, 3506
Pu, H.-Y., Hirotani, K., Mizuno, Y., & Chang, H.-K. 2012, arXiv:1211.1577
Richards, J. L., Hovatta, T., Max-Moerbeck, W., et al. 2014, MNRAS, 438, 3058
Richards, J. L., Max-Moerbeck, W., Pavlidou, V., et al. 2011, ApJS, 194, 29
Sbarrato, T., Ghisellini, G., Maraschi, L., & Colpi, M. 2012, MNRAS, 421, 1764
Sbarrato, T., Ghisellini, G., Tagliaferri, G., et al. 2016, MNRAS, 462, 1542
Scarpa, R., & Falomo, R. 1997, A&A, 325, 109
Shakura, N. I., & Sunyaev, R. A. 1973, A&A, 24, 337
Sikora, M., & Begelman, M. C. 2013, ApJL, 764, L24
Sikora, M., Stasińska, G., Kozieł-Wierzbowska, D., Madejski, G. M., & Asari, N. V. 2013, ApJ, 765, 62
Stickel, M., Padovani, P., Urry, C. M., Fried, J. W., & Kuehr, H. 1991, ApJ, 374, 431
Tan, C., Xue, R., Du, L.-M., et al. 2020, ApJS, 248, 27
Tavecchio, F., Maraschi, L., & Ghisellini, G. 1998, ApJ, 509, 608
Tchekhovskoy, A., Narayan, R., & McKinney, J. C. 2011, MNRAS, 418, L79
Urry, C. M., & Padovani, P. 1995, PASP, 107, 803
Wang, J. M., Ho, L. C., & Staubert, R. 2003, A&A, 409, 887

Wang, J.-M., Staubert, R., & Ho, L. C. 2002, ApJ, 579, 554
Xiao, H., Ouyang, Z., Zhang, L., et al. 2022, ApJ, 925, 40
Xiong, D. R., & Zhang, X. 2014, MNRAS, 441, 3375
Yang, H., Yuan, F., Li, H., et al. 2024, SciA, 10, eadn3544
Yuan, F., & Narayan, R. 2004, ApJ, 612, 724
Yuan, F., & Narayan, R. 2014, ARA&A, 52, 529
Zamaninasab, M., Clausen-Brown, E., Savolainen, T., & Tchekhovskoy, A. 2014, Natur, 510, 126
Zhang, J., Zhang, S.-N., Liang, E.-W., & Sun, X. N. 2014, in 40th COSPAR Scientific Assembly, E1.5–10–14
Zhang, L., Chen, S., Xiao, H., Cai, J., & Fan, J. 2020, ApJ, 897, 10
Zhang, L., Liu, Y., & Fan, J. 2022, ApJ, 935, 4